# Graphene photodetectors with a bandwidth >76 GHz fabricated in a 6" wafer process line

Daniel Schall*, Caroline Porschatis, Martin Otto, Daniel Neumaier

Advanced Microelectronic Center Aachen (AMICA), AMO GmbH, Otto-Blumenthal-Strasse 25, 52064 Aachen Germany

*E-mail: schall@amo.de

**Abstract**

**In recent years, the data traffic has grown exponentially and the forecasts indicate a huge market that could be addressed by communication infrastructure and service providers. However, the processing capacity, space, and energy consumption of the available technology is a serious bottleneck for the exploitation of these markets. Chip-integrated optical communication systems hold the promise of significantly improving these issues related to the current technology. At the moment, the answer to the question which material is best suited for ultrafast chip integrated communication systems is still open. In this manuscript we report on ultrafast graphene photodetectors with a bandwidth of more than 76 GHz well suitable for communication links faster than 100 GBit/s per channel. We extract an upper value of 7.2 ps for the timescale in which the bolometric photoresponse in graphene is generated. The photodetectors were fabricated on 6" silicon-on-insulator wafers in a semiconductor pilot line, demonstrating the scalable fabrication of high-performance graphene based devices.**

Optical communication in general and especially chip integration of optic and electronic components has been considered as a promising way to significantly increase the performance of datalinks in terms of capacity, energy consumption, and costs [1-3]. The current bottleneck that hinders the mass production of chip integrated electro-optic components like modulators and photodetectors is the lack of a material that is compatible to established processing technology of chips while giving high performance devices. The electronic and optic properties of graphene [4], the two dimensional allotrope of carbon, were studied extensively in the last years [5-9]. Moreover, competitive chip-integrated electro-optical devices like electro-optical modulators [10,11], efficient waveguide heaters [12] and ultrafast photodetectors [13-16] were fabricated using graphene as active material. The performance of graphene photodetectors on integrated silicon waveguides in terms of speed and sensitivity has improved significantly in the last years and the gap between the performance of graphene and competing technologies is vanishing [17-21]. The possible monolithic integration on various substrates that are not necessarily crystalline is a key merit that distinguishes graphene from Ge or III/V materials. The bandwidth reported for graphene photodetectors of up to 65 GHz [16] and sensitivity around 0.4 A/W without bias and 1 A/W with bias [22] underline the potential of graphene for chip integrated photodetectors. Besides an excellent device performance the integration in a large scale production environment is essential. However, the introduction of new materials into an existing fabrication process flow and the required adoption as well as development of entirely new process steps are among the most challenging tasks in the fabrication of integrated circuits. So far, graphene photodetectors on silicon waveguides were fabricated on small chips and the transition to wafer scale processing is still missing. Here we report on ultrafast waveguide integrated graphene photodetectors that were fabricated on 6" silicon on insulator (SOI) wafers in a semiconductor pilot line. We developed a wafer-based process flow compatible with graphene, taking the next step towards scalable fabrication of photonic graphene devices.

A schematic of a graphene photodetector integrated on a Si waveguide is shown in figure 1 (a), with the graphene layer placed on top of the waveguide within the evanescent field that surrounds the waveguide, see figure 1 (b). Along the length $L$ marked in the schematic, the light is absorbed and by applying an external bias voltage a photo signal is generated due to the bolometric effect [9]. We utilized the bolometric effect for photo signal generation because this effect does not require an exact location of a doping gradient as required for the photovoltaic or thermoelectric effect [23]. In contrast to that, a bolometric photo current is generated due to a temperature induced resistance variation of a graphene channel that is independent of the exact location of the excitation [9]. We take advantage of the fact that for the bolometric effect the entire graphene channel contributes to the generation of the photo current and circumvent the fabrication constraint related to an exact positioning of a defined doping gradient. Additionally, a bolometric signal generation works well for the expected graphene doping level after the fabrication [9]. Therefore, the detector layout used here represents one of the least complex layouts for waveguide integrated photodetectors and is therefore well suited as a model device for demonstrating the basic feasibility of wafer-scale processing.

The processing started with the definition of the photonic waveguide layer on SOI wafers with 3 µm buried oxide and 220 nm top silicon using an optical I-line stepper. To obtain smooth waveguide edges a resist reflow technique was applied followed by reactive ion etching of the Si. At the location of the graphene photodetector the waveguides had a width of 800 nm. Grating couplers, which are used to couple light from a fiber into the waveguide, were fabricated by high resolution electron beam lithography followed by dry etching. To avoid cracks in the graphene layer, the step edges of the waveguide were smoothened by a layer of spin on glass that was spin coated, cured at 300°C, and coated with 1 nm $Al_2O_3$ by plasma-assisted atomic layer deposition. The total thickness of the cladding on top of the waveguide was 40 nm. At this stage of device fabrication, the optical properties of the photonic structures were measured. All optical and electro-optical measurements were carried out on wafer at room temperature in normal atmosphere. The waveguide and grating coupler losses were determined by transfer length measurements on reference structures. Figure 2 (a) shows that the loss of one grating coupler was 3.1 dB corresponding to 49 % coupling efficiency (2.5 dB for the best couplers), and the loss of the waveguides was 0.36 dB/cm. The waveguides that were intended for the fabrication of the photodetectors were designed symmetrically with the photodetector located in the middle and grating couplers on either end of the waveguide. To determine the optical power available in the middle of the waveguide at the graphene photodetectors we measured the transmission spectrum of these waveguides. The power attenuation at the photodetector is half of the measured fiber-to-fiber insertion loss of the waveguide and grating couplers. In figure 2 (b) an exemplary transmission spectrum is shown with a peak transmission at the design wavelength of 1550 nm and a fiber to fiber insertion loss of 6 dB. The inset in figure 2 (b) shows a histogram of the waveguides before fabrication of the photodetectors at 1550 nm with an average optical loss of 6.3 dB on an entire die. This insertion loss of 6.3 dB corresponds well to the grating coupler and waveguide losses extracted from figure 2 (a) and is dominated by the grating couplers. The optical power level at the graphene photodetector can therefore be calculated by subtracting 3.1 dB from the output power at the optical input fiber.

Subsequently after the optical characterization, commercial graphene grown by CVD on a copper foil (Graphenea S.A.) was coated with 400 nm of PMMA for transfer. The copper foil was etched with

HCl and FeCl$_3$ solutions and transferred to the wafer similar to the method presented in [24]. The graphene was transferred manually to the amorphous surface of the SOI wafer (Al$_2$O$_3$ coated spin on glass), see figure 3 (a) for a photograph of the wafer including the graphene. We note that the automated transfer on wafer-scale is one of the major issues that need to be solved for volume fabrication. Here, we transferred graphene to a single die which already enables extraction of the relevant data. A scanning electron beam image of a typical area around a waveguide shows the crack free deposition of graphene (figure 3 (b) and (c)). The graphene was patterned using optical lithography and O$_2$ dry etching. The contacts were fabricated on top of graphene by optical lithography and sputter deposition. We used AZ 5214E photoresist in the image reversal mode to create a resist profile with an undercut required for the lift-off after the metallization of 20 nm nickel and 100 nm aluminum. Figure 3 (d) shows a photograph of the processed wafer. In figure 3 (e) an optical micrograph of metal pads for photodetectors is shown as well as an SEM image of a photodetector detail after graphene structuring and contact fabrication in figure 3 (f).

The characterization of the photodetectors started with the measurements of the device resistance. The resistance of nine photodetectors measured several times (in total 16 measurements) ranged from 127 Ω to 240 Ω for a bias of 1 V and is plotted in the histogram in figure 4 (a). The device to device variation of the resistance can be explained by a mobility and doping level variation also present in earlier studies, and inhomogeneities in the contact resistance. The change of device resistance between measurements is related to a change in the doping level due to the absorption or desorption of adsorbates on the graphene, as the devices were not encapsulated and measurements were performed under ambient conditions. The measured device resistance of 127 Ω to 240 Ω for a 50 µm wide and 1.8 µm long two terminal device can be explained well by a sheet resistance of order 1 kΩ and a contact resistance of the order of several kΩµm, which is are typical values for graphene devices fabricated using optical lithography [25].

The low frequency response of the photodetectors was measured with a lock-in amplifier at 1 kHz and a modulated laser source at 1550 nm. A photo current was only present in the case of an external bias voltage showing a negative sign with respect to the DC current. Thus the dominant generation mechanism of the photo current can be attributed to the bolometric effect [9]. The sensitivity of a typical detector with a resistance of 130 Ω and a bias voltage of 1V was 1 mA/W (0.13 V/W). This sensitivity was approximately one order of magnitude above the value reported by Freitag et al. for graphene flakes in a normal incidence setup, mainly due to the significantly increased absorption of light in our waveguide configuration [9].

In order to determine the performance at high frequencies, a heterodyne setup according to the schematic shown in the supporting information of reference [16] was used. The optical radio frequency signal was generated by the superposition of two optical signals around 1550 nm, where one laser source was held at a constant wavelength and the other laser was tuned to generate the desired beating frequency. This heterodyne signal was amplified by an erbium doped fiber amplifier to 18 dBm and coupled to the graphene photodetector. The electrical power generated by the photodetector was coupled to a power meter with an impedance of 50 Ω using a 67 GHz GS wafer prober and a 10 cm RF cable. Figure 4 (b) shows an exemplary frequency response for different bias voltages up to 110 GHz of a photodetector with a resistance of 136 Ω at 0.25 V, 138 Ω at 0.5 V, and 140 Ω at 1V bias. This photodetector was selected for detailed analysis because of its low resistance that is close to the 50 Ω of the power meter. The low impedance mismatch led to a high output power which in turn enabled a characterization of the device in a wide variation of the bias voltage. The 3 dB bandwidth for this photodetector including the measurement system was 67 GHz and a dependence of the bandwidth on the bias voltage could not be observed.

The output power at 1 GHz was -49 dBm corresponding to a RF sensitivity of 0.4 mA/W. Taking the impedance mismatch between the power meter and the photodetector into account, the DC and RF sensitivity are in good agreement. In contrast to the bandwidth, the output power depended on the bias voltage and increased by 6 dB for a factor of 2 larger bias voltage. This behavior can be explained by the relation between the bias voltage and the bolometric effect in graphene. If the bias voltage is not too large, the bolometric photo current $I_{ph}$ increases linearly with the bias voltage according to $I_{ph} = V_{bias} / (\beta * T)$ with the bolometric coefficient $\beta = dR/dT$ and the light induced temperature increase $dT$ [9]. With $P = I_{ph}^2/R$, the power of the microwave output signal depends quadratically on the photo current, hence an increase of a factor of two more bias voltage generates 6 dB more electrical output power. Within the achievable measurement resolution the expected 6 dB step can be observed in figure 4 (b).

We measured the frequency response of the nine photodetectors several times at 1V bias (in total 16 measurements) and plotted the histogram of the bandwidth in figure 4 (c), showing that we do not observe a strong variation in the measured bandwidth. In the inset in figure 4 (d) we show exemplarily the frequency response of two photodetectors with a bandwidth of 65 GHz (green curve) and 76 GHz (red curve) representing the lower and upper bandwidth limit. The average bandwidth is approximately a factor of 2 larger than the value of 41 GHz we reported earlier in similar devices [15]. There, the bandwidth of the photodetector was limited by the RC component of the device, with R being the device resistance and C dominated by the capacitance of the contact pads. In this study, we reduced the pad area and thus the contact pad capacitance by approximately a factor of 8, while the device resistance was similar to the values in [15]. However, the measured bandwidth increased only by a factor of 2, which then cannot be explained by an RC limited device characteristic any more. To get further insight into the limitations in the present photodetectors, we measured the frequency response of several devices on the die and plotted the bandwidth versus the resistance of the corresponding photodetector in figure 4 (d). For an RC limitation of the frequency response we would expect a bandwidth variation of a factor of 2 according to $f_{3dB} \propto 1/R$. In figure 4 (b) it is clearly visible that the measured bandwidth of the photodetectors did not depend on the device resistance. We therefore conclude that the measured bandwidth of our photodetectors was not limited by the RC product of the device but limited by the used measurement equipment, which was specified for operation up to 67 GHz (probe tip, cable and bias tee). The real bandwidth of our photodetectors was therefore significantly larger than 76 GHz, which opens the door for their application in data communication links, radar systems, and satellite communication operating in the W-band from 75 to 110 GHz. The measured bandwidth of 76 GHz translates into a response time $\tau = 0.55 / f_{3dB} = 7.2$ ps [26], which shows that the heating and cooling processes associated to the bolometric signal generation in monolayer CVD graphene are taking place at a time scale of 7.2 ps or shorter. This is the first experimental demonstration that the timescale of the bolometric photoresponse in graphene can be in the same order of magnitude like the intrinsic response time of 1.3 to 3 ps for the generation of a photosignal in a doping gradient without external bias due to the thermo-electric effect [22, 27, 28].

To investigate the performance of the detector in a communication system, we integrated the photodetector in a datalink according to the schematic in figure 5 (a). A pseudo random bit pattern generator was used to generate an on-off-keyed bit stream at 12.5 GBit/s. This stream was encoded onto an optical carrier at 1550 nm by a commercial lithium niobate modulator, amplified and coupled to the graphene photodetector. The detected bit stream was amplified and displayed on an oscilloscope. Figure 5 (b) shows the eye diagram of the recorded bit stream for an optical input power of 18 dBm and a bias of 2 V. The clearly reproduced '1' and '0' levels result in an open eye, demonstrating the successful transmission of the data stream. The upper limit of the data rate was defined by the measurement equipment, not the graphene photodetector. According to the Shannon-

Hartley theorem, the measured bandwidth of at least 76 GHz is sufficient for transmission of more than 100 GBit/s in a single channel with the most simple on-off-keying modulation scheme.

In summary, we present for the first time a 6" wafer based fabrication process for opto-electronic graphene devices. The successful monolithic integration of graphene into an existing photonic wafer-level fabrication platform enables the scalable fabrication of graphene photodetectors with a bandwidth larger than 76 GHz. We showed that this monolithic integration and high speed performance is possible on amorphous substrates which cannot be used for the growth of high quality Ge or III/V material layers required for high speed photodetectors. The new bandwidth record is a significant step forward towards the goal of chip integrated high speed optical interconnects with a capacity of more than 100 GBit/s. We were able to demonstrate that the timescale of 7.2 ps of the bolometric photoresponse in graphene is in the same order of magnitude like the intrinsic response times without bias at a doping gradient. Moreover, the demonstrated bandwidth enables the realization of Si-integrated microwave photonic devices operating in the W-band.

**Acknowledgement**

We acknowledge funding by the European Commission within the project "Graphene Flagship" (contract no. 696656). We acknowledge electron beam lithography by Jens Bolten, Georg Götz, and Thorsten Wahlbrink as well as optical lithography by Holger Lerch. We thank R. Negra for providing measurement equipment.

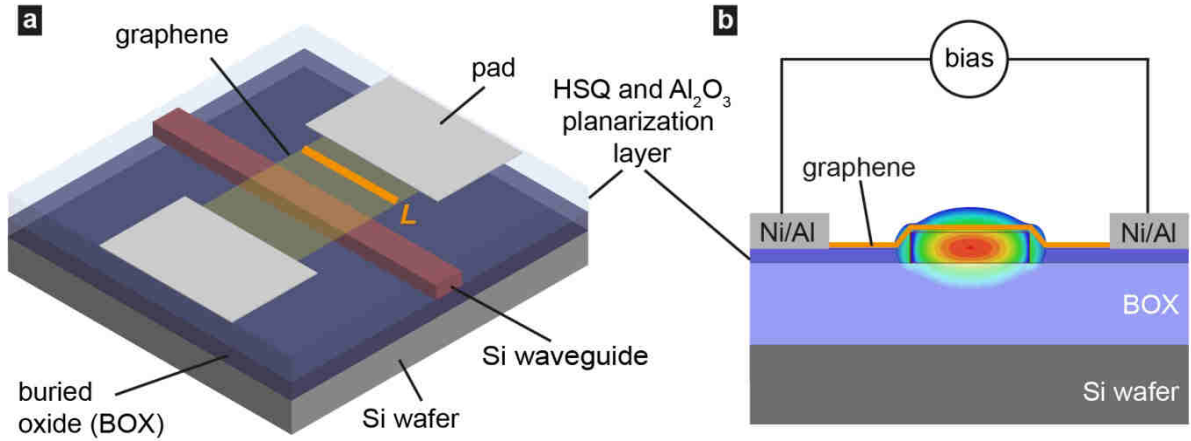

**Figure 1 (a)** Schematic illustration of an integrated graphene photodetector on a Si waveguide. Along the length *L* of the device the light is absorbed. **(b)** Schematic of the device showing the cross section through a graphene photodetector. The evanescent field that surrounds the waveguide is illustrated by the red and greed mode field, the graphene is illustrated as the white line within the evanescent field.

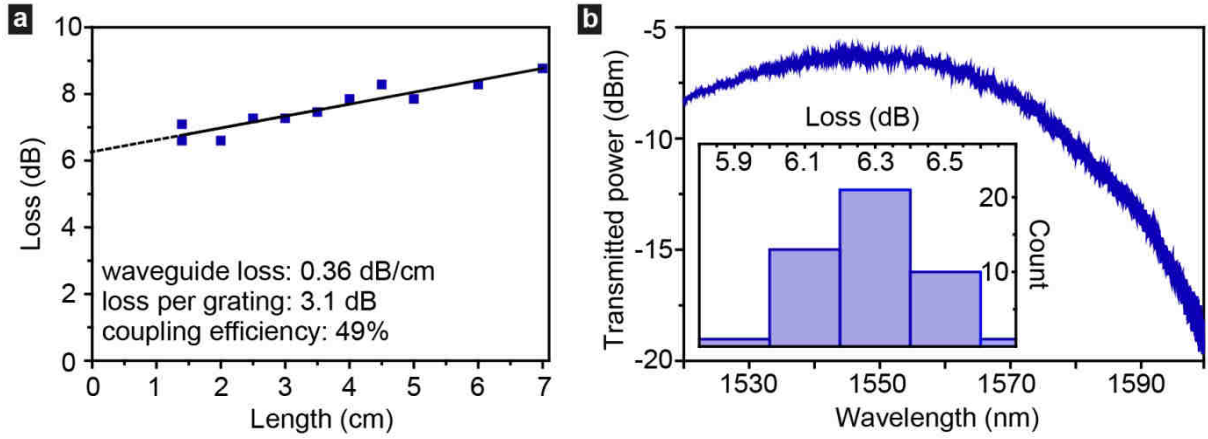

**Figure 2 (a)** Transfer length measurements of waveguides. The slope gives the waveguide loss of 0.36 dB/cm and the intersection with the y-axis the grating coupler loss of 6.2 dB for the two couplers together. **(b)** Optical transmission spectrum of a typical waveguide on which photodetectors were fabricated on. The peak transmission is at 1550 nm. For this wavelength the total fibre-to-fibre loss of the waveguides on one die is shown in the histogram.

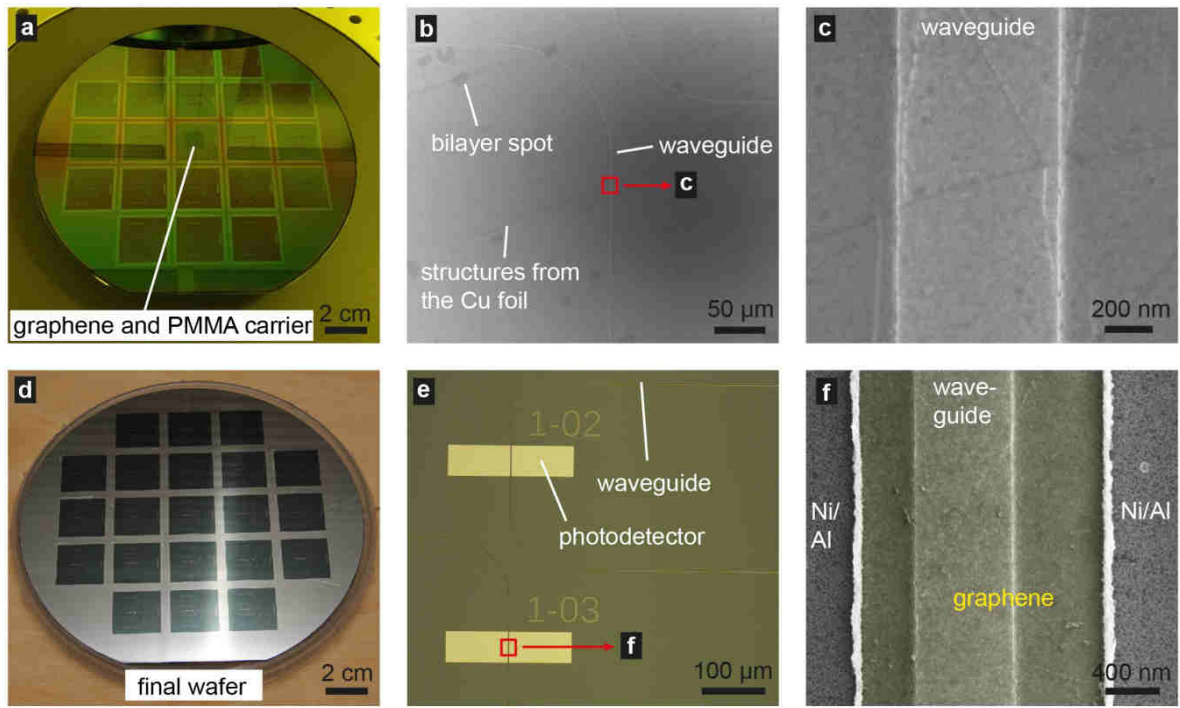

**Figure 3 (a)** 6" SOI wafer after fabrication of the photonic layer including the anti-cracking layer. In the center die the transferred graphene including the PMMA carrier is visible as a gray square. **(b)** SEM image of the transferred graphene after removing the PMMA carrier. The Si waveguides were oriented from left to right and bent into the up to down direction where the detector was located. The visible dark spots are bilayer areas that form during the graphene growth by CVD on copper foil. **(c)** Detailed SEM image of the transferred graphene that conformably follows the surface over the waveguide without cracks. The wrinkles are typical for CVD grown graphene. **(d)** Photo of the wafer after the final metallization. **(e)** Optical micrograph of the final metal pads. As in (b) the waveguides run from left to right and bend into the up down direction. The pads are fabricated in the middle of the waveguides. **(f)** Detailed SEM image of a fabricated photodetector showing the edge of the two metal pads as well as the waveguide in the center. Graphene covers the entire area between the contact pads.

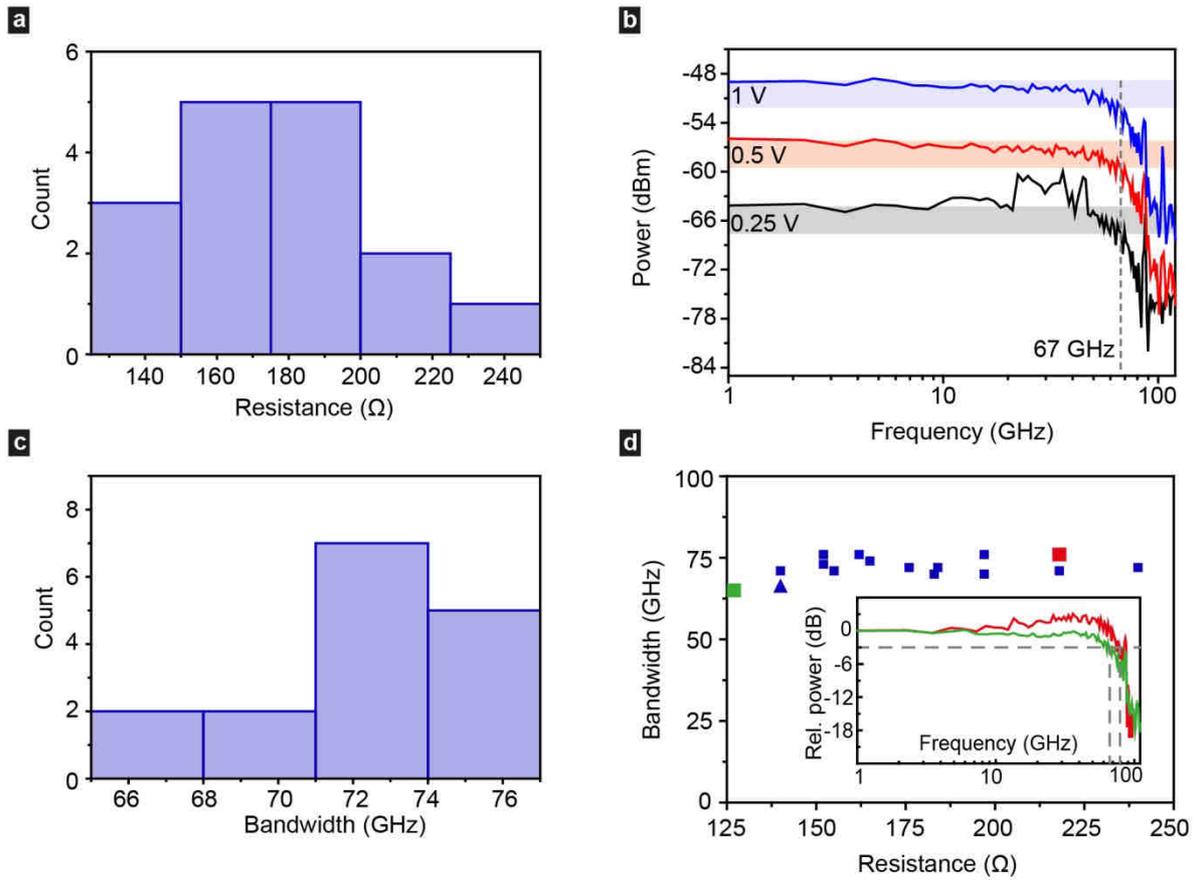

**Figure 4 (a)** Histogram of the device resistance. **(b)** Frequency response of a typical photodetector for 0.25, 0.5 and 1 V bias voltage. **(c)** Histogram of the photodetector bandwidth. **(d)** Photodetector bandwidth and corresponding device resistance. The inset shows frequency responses from the lower (green $f_{3dB}$ = 65 GHz) and upper limit (red $f_{3dB}$ = 76 GHz) of the measured bandwidths. The green and red data points in the main figure correspond to the colors of the frequency responses in the inset. The triangle represents the detector characterized in detail in figure (b).

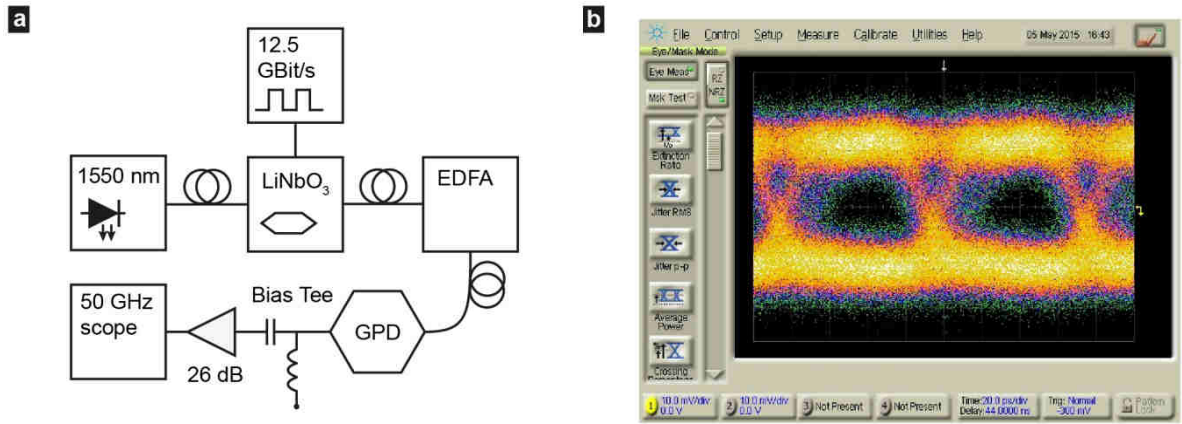

**Figure 5 (a)** 12.5 GBit/s data link. The data sequence is coupled from a PRBS generator to an electro-optical LiNbO$_3$ modulator that generates the optical signal. The optical data stream is amplified and coupled to the graphene photodetector. The detected bit sequence is amplified and displayed in a sampling oscilloscope. **(b)** Recorded eye diagram showing the successful transmission of a data stream at 12.5 GBit/s.